\documentclass[fleqn,usenatbib]{mnras}

\usepackage[T1]{fontenc}
\usepackage{ae,aecompl}

\usepackage{amsmath}	
\usepackage{amssymb}	
\usepackage{graphicx}
\usepackage{hyperref}

\newcommand{\angstrom}{\mbox{\normalfont\AA}}

\title[X-ray source in LyC candidate Tol 0440-381]{Rapid turn-on of a luminous X-ray source in the candidate Lyman continuum emitting galaxy Tol 0440-381}

\author[Kaaret, Bluem, Prestwich]{
P. Kaaret$^{1}$ \thanks{E-mail: philip-kaaret@uiowa.edu},
J. Bluem$^{1}$
A.H. Prestwich$^{2}$,
\\
$^{1}$Department of Physics and Astronomy, University of Iowa, Iowa City, IA 52245, USA\\
$^{2}$Harvard-Smithsonian Center for Astrophysics, Cambridge, MA, 02318, USA\\
}

\date{Accepted XXX. Received YYY; in original form ZZZ}

\pubyear{2021}

\begin{document}
\label{firstpage}
\pagerange{\pageref{firstpage}--\pageref{lastpage}}
\maketitle

\begin{abstract}

{\it Chandra} observations of the nearby, candidate Lyman-continuum (LyC) emitting galaxy Tol 0440-381 show brightening of an X-ray source by at least a factor of 4 to a luminosity of $1.6 \times 10^{40} \rm \, erg \, s^{-1}$ over 3.8~days. The X-ray emission likely arises from either a low-luminosity AGN or an ultraluminous X-ray source. The properties of the X-ray source are similar to those found in Haro~11 and Tololo~1247-232, the only other LyC-emitting galaxies that have been resolved in X-rays. All three galaxies host luminous, variable, and hard spectrum X-ray sources that are likely accretion-powered. Accretion onto compact objects produces powerful outflows and ionizing radiation that could help enable LyC escape.

\end{abstract}


\begin{keywords}
galaxies: star formation -- galaxies: individual: Tol 0440-381 -- X-rays: galaxies -- X-rays: binaries
\end{keywords}


\section{Introduction}

The escape of Lyman continuum (LyC) radiation from early galaxies was essential to reionize the universe at high redshifts. However, LyC is strongly absorbed by interstellar material to such a degree that most nearby galaxies produce no detectable LyC emission. Mechanical feedback has been suggested as a mechanism to enable LyC escape by creating channels with low optical depth in the interstellar medium \citep{Wofford2013,Orsi2012}. Stellar winds and supernovae ejecta may provide the needed feedback \citep{Tenorio1999,Hayes2010}. Outflows from accreting compact objects provide another potential source of mechanical power of comparable magnitude \citep{Prestwich2015}. Determining whether or not LyC emitting galaxies commonly host accretion-powered X-ray sources should provide insight to whether outflows from accreting objects help enable LyC escape.

The few nearby galaxies with LyC escape serve as local laboratories that can be resolved with current X-ray and optical instrumentation. The LyC escape fraction is defined as the ratio of the observed (escaping) luminosity at 900~\AA\ in the rest frame of the galaxy to the luminosity at the same wavelength produced by the stars before attenuation by gas and dust. We define a LyC emitter as a galaxy from which non-zero LyC emission has been detected and require no minimum value for the escape fraction. There are only three nearby ($z < 0.1$) galaxies that are confirmed or candidate LyC emitters: Haro~11, Tololo~1247-232, and Tololo~0440-381.\footnote{LyC detection from Mrk~54 was reported by \citet{Leitherer2016}. However, \citet{Chisholm2017} place an upper limit on its LyC escape fraction of 0.16 per cent and do not consider the source to be a LyC emitter.}  All three are compact galaxies with high star formation rates. Table~\ref{LyC_table} presents some basic properties of the three galaxies.

\begin{table*}
\caption{Properties of Lyman continuum-emitting galaxies.}
\begin{tabular}{lccccccc}
\hline
Name         & $\log(M_{*})$ 
                    & SFR & log(O/H)+12 
                                & $f_{\rm esc}$       & Galaxy $L_{\rm X}$ 
                                                                            & Point $L_{\rm X}$ 
                                                                                                  & $\Gamma$ \\
             & $\log(M_{\odot})$ 
                    & $M_{\odot} {\rm yr}^{-1}$ &
                                & per cent            & $10^{40} \rm \, erg \, s^{-1}$
                                                              & $10^{40} \rm \, erg \, s^{-1}$ &  \\  \hline
Haro 11      & 10.1 & 26  & 8.1 & $3.3 \pm 0.7$       &  16   & $4.2\pm 2.6$        & 1.5 \\
             &      &     &     &                     &       & $3.7 \pm 1.2$       & 1.8 \\
Tol 1247-232 &  9.7 & 16  & 8.1 & $0.4^{+1.9}_{-1.7}$ &  18   & $9 \pm 2$           & 1.6 \\
Tol 0440-381 & 10.0 & 13  & 8.2 & $1.9^{+1.1}_{-1.0}$ &   3   & $1.6^{+1.0}_{-0.7}$ & 1.6 \\ \hline
\end{tabular} 
\newline
\begin{flushleft}
Note: Columns: galaxy name, stellar mass, star formation rate, oxygen to hydrogen abundance ratio, LyC escape fraction, average galactic X-ray luminosity, maximum point source X-ray luminosity, point source photon index. There are two lines for Haro 11 for the two X-ray emitting regions. Stellar masses, metallicities, and SFRs for Haro 11 and Tol1247 are from \citet{Chisholm2017}. Escape fractions are from \citet{Chisholm2017}, except \citet{Leitet2011} for Haro 11. X-ray quantities are from \citet{Gross2021} for Haro 11 and from \citet{Kaaret2017} for Tol1247. SFR and X-ray quantities for Tol0440 are from this work and the X-ray luminosity is the average for observations A and B from Table~\ref{xray_sources}.
\end{flushleft}
\label{LyC_table}
\end{table*}

Haro~11 is the first galaxy from which LyC emission was detected, using the {\it Far Ultraviolet Spectroscopic Explorer} (FUSE) \citep{Bergvall2006}. The escape fraction was measured to be between 4 to 10 per cent. Analyzing the same data set with different reduction techniques, \citet{Grimes2007} placed an upper limit of 2 per cent on the escape fraction. \citet{Leitet2011} subsequently performed an analysis with an improved background estimation technique and detected LyC emission with an escape fraction of $3.3 \pm 0.7$ per cent.

LyC emission was first reported from Tol 1247-232 (hereafter Tol1247) with an escape fraction of $2.4^{+0.9}_{-0.8}$ per cent using FUSE data \citep{Leitet2013}. Subsequently, \citet{Leitherer2016} reported an escape fraction of $4.5 \pm 1.2$ per cent using data from the Cosmic Origins Spectrograph (COS) on board the {\it Hubble Space Telescope} (HST). \citet{Chisholm2017} performed an analysis of the same observations, but including only exposures obtained while HST was in the Earth's shadow to reduce the effect of geocoronal Lyman series emission and scattered light. They measured an escape fraction of $0.40^{+0.19}_{-0.17}$ per cent and regard the apparent detection in the shadow data as an upper limit. Hence, Tol1247 is a candidate LyC emitter.

Tol~0440-381 (hereafter Tol0440) is an {H\sc{ii}} galaxy with possible Wolf Rayet features and a stellar mass of $10^{10} M_{\odot}$ \citep{Leitet2013}. Data from HST/COS were used to study LyC emission from Tol~0440-381. \citet{Leitherer2016} measured an escape fraction of $7.1 \pm 1.1$ per cent, but interpreted it as an upper limit due to possible contamination by geocoronal Lyman series lines. \citet{Chisholm2017} analysed the same observations, using only exposures in orbit shadow. They measured an escape fraction of $1.9^{+1.1}_{-1.0}$ per cent. They conservatively interpret the apparent detection in the shadow data as an upper limit. Hence, Tol0440 is a candidate LyC emitter.

Haro 11 and Tol1247 are known X-ray emitters. Haro 11 has a total X-ray luminosity $\sim 10^{41} \rm \, erg \, s^{-1}$ and contains two X-ray emitting regions, each coincident with a star-forming region \citep{Grimes2007,Prestwich2015}. The emission from each region is resolved into two sources plus extended emission, with the individual sources showing variability at the level of $\sim 4 \times 10^{40} \rm \, erg \, s^{-1}$ on time scales of years \citep{Gross2021}. The emission from the brighter region is quite hard with a photon index $\Gamma \sim 1.5$, while that from the dimmer region is moderately hard with $\Gamma \sim 1.8$. X-ray emission from Tol 1247-232 was first detected at $L_X = 2.3 \times \sim 10^{41} \rm \, erg \, s^{-1}$ with {\it XMM-Newton} which lacks the spatial resolution needed to the resolve the galaxy \citep{Rosa2009}. A subsequent observation with the {\it Chandra X-ray Observatory} revealed a single unresolved source with $L_X \sim 9 \times 10^{40} \rm \, erg \, s^{-1}$ and extended emission \citep{Kaaret2017tol}. The total X-ray flux varied by a factor of $\sim 2$ between two observations separated by 11 years. The spectrum of the unresolved source detected with {\it Chandra} is hard, $\Gamma \sim 1.6$.

Modelling of the star-forming regions in these two galaxies indicates the total mechanical power generated by stellar winds and supernovae is comparable to the luminosity of the X-ray sources \citep{Prestwich2015,Kaaret2017tol}. The outflow power of an accreting object is often comparable and sometimes exceed its radiative luminosity \citep{Gallo2005}. Therefore, the outflows from the X-ray sources may help enable Lyman escape. Hence, it is of interest to determine whether other LyC emitting galaxies host accretion-powered X-ray sources.

We obtained X-ray observations of Tol0440 with {\it Chandra}. We describe the observations and our analysis in section~\ref{sec:obs}, then our results and interpretation in section~\ref{sec:results}. We adopt a distance to Tol0440 of 167~Mpc \citep{Leitherer2016}.

\begin{figure*}[ht]
\centerline{\includegraphics[width=5.5in]{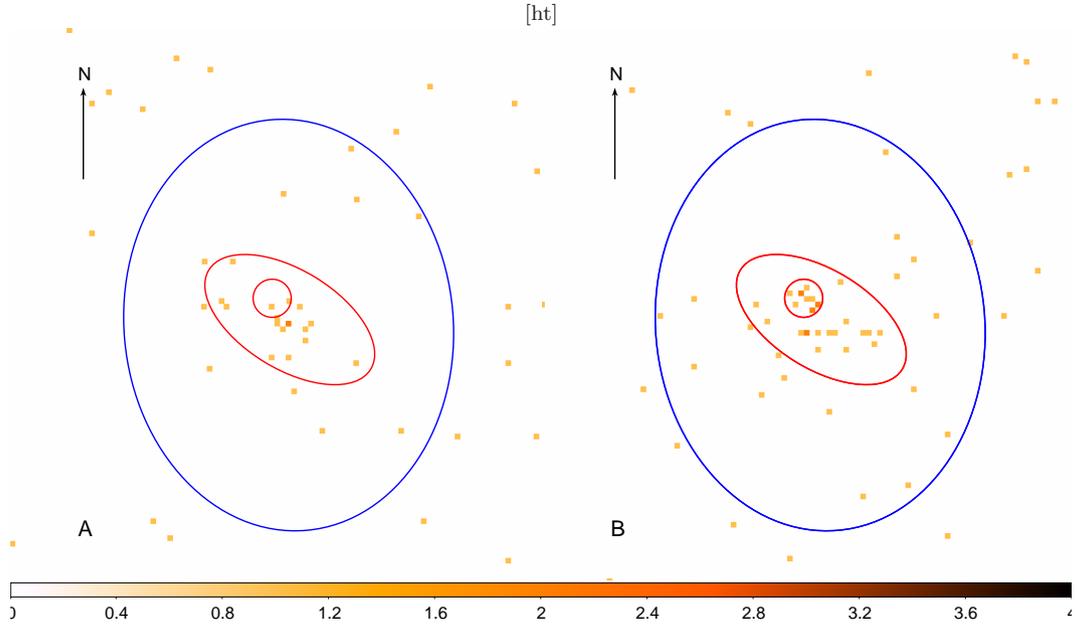}}
\caption{X-ray images of Tol 0440-381. The left panel shows observation A and the right panel shows observation B. Both images show counts in the 0.5-8 keV band. The pixels are $0.246\arcsec$ or 199~pc at the 167~Mpc distance of Tol 0440-381. The large blue ellipse shows the extent of the galaxy in the NUV as measured with GALEX. The red ellipse encloses the total X-ray emission. The red circle encloses the transient X-ray source present in observation B and absent in observation A. The black arrow points North and has a length of $4\arcsec$. The exposure for B is $1.26\times$ longer than that for A, so it contains more counts due to background and extended emission.}
\label{xray_images}
\end{figure*}

\section{Observations and Analysis}\label{observations}
\label{sec:obs}

Our approved 64~ks of {\it Chandra} observing time was divided into two observations for operational reasons. The first observation had an exposure of 27.70~ks and began at 2021-02-16 02:31:16 UTC (ObsID 24778, hereafter observation A). The second had an exposure of 34.79 ks and began 3.8~days later at 2021-02-19 22:21:4902:26:01 UTC (ObsID 22486, hereafter observation B).

We used CIAO version 4.13 and CALDB version 4.9.4 for our analysis. We aligned the observations using X-ray sources detected in both using the CIAO tool {\tt wavdetect}. We included only sources on the S3 chip where the target galaxy is located. A total of 18 matching sources were found. We used the CIAO tool {\tt wcs\_match} to calculate the relative astrometry and allowed only translations. The root mean square residual of the source positions between the two observations was $0.34\arcsec$ after the alignment.

Figure~\ref{xray_images} shows X-ray images in the 0.5-8~keV band from the two observations using $0.246\arcsec$ pixels. There is extended X-ray emission coincident with Tol0440 in both observations. Running {\tt wavdetect} on a merged image of the two observations in the 0.5-8~keV band, we find an extent of $4.1\arcsec \times 2.2\arcsec$ as shown by the red ellipse in the figure. There is emission consistent with an unresolved source in observation B (within the red circle) that is not present in observation A. There are 11 counts total (without any background subtraction) in observation B within a circle with a radius of $0.833\arcsec$ enclosing 90 per cent of the point spread function at 1.0~keV. Scaling from the level of diffuse emission in observation A, only 1.3 counts are expected. The chance probability of occurrence of 11 counts within a radius of $0.833\arcsec$ assuming a Poisson distribution and allowing for 12.9 trials equal to the ratio of the areas of the ellipse and circle is $2.4 \times 10^{-6}$. Hence, the source is significantly detected in observation B above the extended emission as measured in observation A. The events from the point source are roughly evenly distributed in time over observations B. The ratio of counts for the point source above/below 2~keV is 6/5. This suggests a hard spectrum, with photon index $\Gamma \sim 1.6$, but is consistent with spectra as soft as $\Gamma \sim 2.2$.

Table~\ref{xray_sources} provides background-subtracted count rates and observed and unabsorbed fluxes for X-rays extracted from the two regions for the two observations. The columns labelled `Galaxy' give the total count rates and fluxes for the $4.1\arcsec \times 2.2\arcsec$ ellipse shown in red in Fig.~\ref{xray_images} while those labelled `Point source' gives values for the $0.833\arcsec$ red circle. The counts and fluxes for `Galaxy' include the `Point source'. All values are calculated for the 0.5-8~keV band. Background was estimated using an annulus centered on Tol0440 with an inner radius of $15\arcsec$ and an outer radius of $25\arcsec$. Observed fluxes and fluxes corrected for absorption within the Milky Way were calculated using an absorbed power-law model with $\Gamma = 2$ and $N_{\rm H} = 1.48 \times 10^{20} \rm \, cm^{-2}$. X-ray luminosities ($L_{\rm X}$) were calculated from the absorption-corrected fluxes assuming a distance of 167~Mpc and isotropic emission. Using $\Gamma = 1.6$ for the point source in observation B leads to a 6 per cent increase in the unabsorbed model flux.

Table~\ref{LyC_table} provides information about the X-ray emission from the three galaxies. We include the average X-ray luminosity of each galaxy. The X-ray point sources are characterized using the largest observed change in X-ray luminosity within each X-ray bright region as an estimate of the maximum point source luminosity along with the corresponding inferred photon index for an absorbed power-law spectrum. Haro 11 contains two X-ray bright regions and there is evidence for multiple sources in each \citep{Gross2021}.

For comparison with the X-ray properties, we estimated the infrared (IR) and near ultraviolet (NUV) luminosities of Tol0440. The blue ellipse in Fig.~\ref{xray_images} was chosen to contain the NUV emission (central wavelength of 231~nm) from the galaxy detected with the {\it Galaxy Evolution Explorer} (GALEX). The position angle of the ellipse is $5\degr$ East of North in agreement with the value from Hyperleda. The extent of the ellipse is $9\arcsec \times 7.2\arcsec$ which is smaller than the extent in Hyperleda and is likely an overestimate due to the $6.0\arcsec$ angular resolution of GALEX \citep{Morrissey2005}. We find a background-subtracted count rate of $25.4 \rm \, s^{-1}$ for Tol0440. This gives an NUV flux density of $5.23 \times 10^{-15} \rm \, erg \, cm^{-2} \, s^{-1} \angstrom^{-1}$ using the conversion factor from the GALEX Observer's Guide. After extinction correction using the \citet{Cardelli1989} extinction curve and a reddening, E(B-V) = 0.0159, the NUV luminosity is then $L_{\rm NUV} = 4.6 \times 10^{43} \rm \, erg \, s^{-1}$.

We calculated the IR luminosity following \citet{Basu2013} and \citet{Brorby2016}. The Wide-field Infrared Survey Explorer (WISE) $22 \mu m$ (band 4) magnitude of $6.164 \pm 0.045$ was converted to a monochromatic luminosity following \citep{Wright2012}. That was then converted to an 8-1000 micron band luminosity using the spectral energy density templates of \citet{Chary2001}. We estimate the IR luminosity of Tol0440 to be $L_{\rm IR} = 1.3 \times 10^{43} \rm \, erg \, s^{-1}$. The ratio of the IR to the UV luminosity agrees well with the correlation with $L_{\rm IR}$ found by \citet{Bell2003}, see also \citet{Iglesias2006}.

Using these IR and NUV luminosities, we estimate $\rm SFR_{IR} = 7.0 \, M_{\odot} \, yr^{-1}$ and $\rm SFR_{NUV} = 5.5 \, M_{\odot} \, yr^{-1}$ using the relations in \citet{Mineo2012} which derive from \citet{Iglesias2006} and assume a Salpeter initial mass function (IMF) from $0.1-100 \, M_{\odot}$. Because Tol0440 is a starburst, the IR emission from active star formation likely dominates that from aged stars  \citep{Hirashita2003}, so the total SFR is the sum of the IR and NUV SFRs, $\rm SFR = 12.5 \, M_{\odot} \, yr^{-1}$. \citet{Leitherer2016} find a SFR of $3.2 \rm \, M_{\odot} \, yr^{-1}$ using UV spectra obtained with HST/COS. Their SFR is for the area of the galaxy viewed through the COS primary science aperture. Scaling by the ratio of the area of extended X-ray emission to the COS aperture ($2.5\arcsec$ diameter) increases the SFR to $4.6 \, M_{\odot} \rm \, yr^{-1}$ in reasonable agreement with our $\rm SFR_{NUV}$. \citet{Leitet2011} quote $\rm SFR = 7.7 \, M_{\odot} \, yr^{-1}$. They state that the SFR for their sample of galaxies was derived by summing the IR and UV SFRs with the IR from the the Infrared Astronomical Satellite (IRAS) and the UV from International Ultraviolet Explorer. However, Tol0440 is detected in only one of the four IRAS bands, which might have caused some issue with the IR contribution. Their value is in reasonable agreement with our $\rm SFR_{NUV}$.



\begin{table}
\caption{X-ray count rates, fluxes, and luminosities.}
\begin{center}
\begin{tabular}{lcccc}
\hline
            & \multicolumn{2}{c}{Galaxy}                 & \multicolumn{2}{c}{Point source} \\
Observation & A                   & B                    & A       & B \\  \hline
Rate        & $5.7_{-2.2}^{+2.8}$ & $7.1_{-2.2}^{+2.8}$  & $<2.0$  & $3.4_{-1.6}^{+2.1}$ \\
Flux        & $8.0_{-3.1}^{+4.0}$ & $10.0_{-3.1}^{+3.9}$ & $<2.8$  & $4.8^{+3.0}_{-2.2}$ \\
Unabs       & $8.2_{-2.3}^{+4.1}$ & $10.3_{-3.2}^{+3.9}$ & $<2.8$  & $4.9^{+3.1}_{-2.2}$ \\ 
$L_{\rm X}$ & $2.7_{-0.8}^{+1.4}$ & $3.4_{-1.1}^{+1.3}$  & $<0.9$  & $1.6^{+1.0}_{-0.7}$ \\ \hline
\end{tabular} 
\newline
\end{center}
Note: Count rates (Rate) are background subtracted and in units of $10^{-4} \rm \, s^{-1}$. Observed fluxes (Flux) and fluxes corrected for absorption within the Milky Way (Unabs) are in units of $10^{-15} \rm \, erg \, cm^2 \, s^{-1}$ and were calculated using an absorbed power-law model with $\Gamma = 2$ and $N_{\rm H} = 1.48 \times 10^{20} \rm \, cm^{-2}$. X-ray luminosities ($L_{\rm X}$) are in units of $10^{40} \rm \, erg \, s^{-1}$ and were calculated from the absorption-corrected fluxes assuming a distance of 167~Mpc. All are for the 0.5-8~keV band.
\label{xray_sources}
\end{table}

\section{Results and Discussion}\label{results}
\label{sec:results}

The {\it Chandra} images, Fig.~\ref{xray_images}, show that Tol0440 contains a variable X-ray source that reaches an intrinsic luminosity of $1.6_{-0.7}^{+1.0} \times 10^{40} \rm \, erg \, s^{-1}$, see Table~\ref{xray_sources}. The variability on timescales of days clearly indicates a single source. The source is mostly likely powered by an accreting compact object.

The emission from the variable source could be due to a low-luminosity AGN (LLAGN). The Type 1 Seyfert galaxy NGC~4395 has shown intraday variability with changes in X-ray luminosity $\sim 1 \times 10^{40} \rm \, erg \, s^{-1}$ \citep{Cameron2012,Kaaret2009} and a hard spectrum, similar to what we observe from Tol0440. The emission line measurements of \citet{Kehrig2004} place Tol0440 within the starforming region of the Baldwin-Phillips-Terlevich (BPT) diagram \citep{Baldwin1981}. However, a minor AGN contribution cannot be excluded. We note that the emission line measurements are for the whole galaxy. Spatially-resolved emission line measurements of the regions surrounding the X-ray sources in Haro~11 lie at the border of the composite AGN/starforming region of the BPT diagram, while the full galaxy lies well within the starforming region \citep{Gross2021}. Spatially-resolved spectroscopy of Tol0440 would be important to better characterize the optical emission near the luminous X-ray source.

The accretion flow for a low-luminosity active galactic nucleus (LLAGN) would have a very low Eddington ratio. Such an accretion flow would produce a hard X-ray spectrum, as inferred for the point source during observation B. In that regime, the accretion would produce an outflow with mechanical energy comparable to or larger than the X-ray luminosity \citep{Ho2008}. Accretion onto an intermediate mass black hole with a mass $\gtrsim 1000 \, M_{\sun}$ would also be sub-Eddington and produce hard X-ray emission and a powerful outflow. Such objects may have formed at early times from minihalos \citep{Madau2004} or recently via dynamical interactions in star clusters \citep{Portegies2004}.

The variable X-ray source may, instead, be an X-ray binary that would be classified as an ultraluminous X-ray source (ULX), for a review, see \citet{Kaaret2017}. The UV spectra of Tol0440 are well matched by a synthetic spectrum for a continuously forming stellar population with an age of 20~Myr \citep{Leitherer2016}. This age allows for the formation of X-ray binaries in which the compact object is either a stellar-mass black hole or a neutron star. In this case, the accretion would be highly super-Eddington and a strong outflow would be expected. 


The total X-ray luminosity of the galaxy varies from 2.7 to 3.4 with an average of $3.1_{-0.7}^{+1.0} \times 10^{40} \rm \, erg \, s^{-1}$, see Table~\ref{xray_sources}. The X-ray luminosity of star-forming galaxies is correlated with their SFR. Using the SFR estimated in the previous section, the metallicity in Table~\ref{LyC_table}, and the relations between SFR and X-ray luminosity from \citet{Brorby2016}, see also \citet{Lehmer2021}, the X-ray luminosity of Tol0440 should be $7.5 \times 10^{40} \rm \, erg \, s^{-1}$. The average X-ray luminosity of the whole galaxy is consistent with the relation, given the 0.34~dex dispersion of the relation and the error on the luminosity, but lies at the low edge. Hence, the total X-ray luminosity of Tol0440 is at or perhaps below the value expected based on the its SFR and metallicity. This is in contrast to the suggestion of \citet{Bluem2019} that X-ray luminosity above that inferred based on the SFR is important to enable LyC escape.

The key factor may, instead, be the presence of luminous, hard-spectrum X-ray sources. All three nearby galaxies with reported LyC emission host at least one luminous and variable X-ray source with a hard spectrum, see Table~\ref{LyC_table}. The variability of the X-ray sources suggests that they are accretion powered. Accretion-powered sources are classified as in the hard spectral state when $1.4 < \Gamma < 2.1$ \citep{Remillard2006}. All of the luminous X-ray sources in the three galaxies are consistent with being in the hard state. The hard state is associated with presence of quasi-steady radio jets, indicating relativistic outflows. The outflow mechanical power while in the hard state is comparable to and sometimes exceeds the X-ray luminosity \citep{Gallo2005,Justham2012,Pakull2010}. Therefore, the X-ray sources likely produce outflows with mechanical power equal to or greater than their X-ray luminosities which exceed $10^{40} \rm \, erg \, s^{-1}$. This is comparable to the mechanical power estimated from supernovae and stellar winds given the SFR within the galaxies and indicates that the accretion driven outflows should be energetically important \citep{Prestwich2015}.

LyC escape can occur via ionized or low-density channels that provide a path for the LyC radiation produced within the star forming regions to pass through the surrounding interstellar medium (ISM) and exit the host galaxy. Luminous accreting sources can help photoionize the ISM. Indeed, \citep{Micheva2018} find an extremely high ionization parameter in Tol1247 that could be explained by the accreting X-ray source. Accretion powered outflows could entrain matter and create an empty channel or could deposit energy and help ionize the ISM. The outflows would be well-collimated near their origin, but can produce bubbles with wider opening angles \citep{Gallo2005} similar to those produced by the jets of active galactic nuclei \citep{Fabian2012}.

Continued X-ray monitoring of Tol~0440-381 is important to understand the temporal behaviour of the X-ray source and to obtain better constraints on the X-ray spectrum. High spatial resolution radio interferometric observations of Tol0440 leading to the detection of a compact radio core coincident with the X-ray emission could provide evidence for an LLAGN or a sub-Eddington intermediate mass black hole or a means to refute that possibility. High-resolution, multiband optical/UV imaging of Tol0440 would be useful to understand the spatial relation of the luminous X-ray source to the star formation activity, particularly the locations of star clusters, and to search for evidence of outflows associated with the X-ray source.

\section*{Acknowledgements}

We thank the referee for comments that improved the manuscript. Support for this work was provided by the National Aeronautics and Space Administration through Chandra Award Number GO0-21075X issued by the Chandra X-ray Observatory Center, which is operated by the Smithsonian Astrophysical Observatory for and on behalf of the National Aeronautics Space Administration under contract NAS8-03060. We acknowledge the use of the HyperLeda database. This research has made use of the NASA/IPAC Infrared Science Archive, which is operated by the Jet Propulsion Laboratory, California Institute of Technology, under contract with the National Aeronautics and Space Administration. 


\section*{Data availability}

The data used in this paper will be made available in the Chandra data archive under ObsIDs 24778 and 22486, \url{https://asc.harvard.edu/cda/}.







\bsp	
\label{lastpage}
\end{document}